# Ly-alpha polarimeter design for CLASP rocket experiment

H. Watanabe[*a], N. Narukage[b], M. Kubo[b], R. Ishikawa[b], T. Bando[b], R. Kano[b],
S. Tsuneta[b], K. Kobayashi[c], K. Ichimoto[a,] J. Trujillo-Bueno[d]

[a]Kwasan and Hida Observatories, Kyoto University, Yamashina-ku, Kyoto, Japan, 607-8417;
[b]National Astronomical Observatory of Japan, 2-21-1 Osawa, Mitaka, Tokyo, Japan 181-8588;
[c]Center for Space Plasma and Aeronomic Research, The University of Alabama in Huntsville, Huntsville, AL, USA 35899
[d]Instituto de Astrofísica de Canarias, Observatorio del Teide, C/Vía Láctea s/n La Laguna, Tenerife, Spain ES 38200

**ABSTRACT**

A sounding-rocket program called the Chromospheric Lyman-Alpha Spectro-Polarimeter (CLASP) is proposed to be launched in the summer of 2014. CLASP will observe the solar chromosphere in Ly-alpha (121.567 nm), aiming to detect the linear polarization signal produced by scattering processes and the Hanle effect for the first time. The polarimeter of CLASP consists of a rotating half-waveplate, a beam splitter, and a polarization analyzer. Magnesium Fluoride ($MgF_2$) is used for these optical components, because $MgF_2$ exhibits birefringent property and high transparency at ultraviolet wavelength.

The development and comprehensive testing program of the optical components of the polarimeter is underway using the synchrotron beamline at the Ultraviolet Synchrotron Orbital Radiation Facility (UVSOR). The first objective is deriving the optical constants of $MgF_2$ by the measurement of the reflectance and transmittance against oblique incident angles for the s-polarized and the p-polarized light. The ordinary refractive index and extinction coefficient along the ordinary and extraordinary axes are derived with a least-square fitting in such a way that the reflectance and transmittance satisfy the Kramers-Krönig relation. The reflection at the Brewster's Angle of $MgF_2$ plate is confirmed to become a good polarization analyzer at Ly-alpha. The second objective is the retardation measurement of a zeroth-order waveplate made of $MgF_2$. The retardation of a waveplate is determined by observing the modulation amplitude that comes out of a waveplate and a polarization analyzer. We tested a waveplate with the thickness difference of 14.57 um. The 14.57 um waveplate worked as a half-waveplate at 121.74 nm. We derived that a waveplate with the thickness difference of 15.71 um will work as a half-waveplate at Ly-alpha wavelength.

We developed a prototype of CLASP polarimeter using the $MgF_2$ half-waveplate and polarization analyzers, and succeeded in obtaining the modulation patterns that are consistent with the theoretical prediction. We confirm that the performance of the prototype is optimized for measuring linear polarization signal with the least effect of the crosstalk from the circular polarization.

## 1. INTRODUCTION

Development work for a proposed joint (Japan, U.S., and Spain) rocket experiment called the Chromospheric Lyman-Alpha Spectro-Polarimeter (CLASP) was started in 2009. CLASP measures the linear polarization profiles in the solar chromosphere and the transition layer with Ly-alpha wavelength (121.567 nm). Ly-alpha is the hydrogen line emitted when an electron transits from the level of principal quantum number $n = 2$ to the $n = 1$ level, and is one of the most intense emission lines in the vacuum-ultraviolet range. In principle, in the presence of magnetic field, the Zeeman splitting of degenerate atomic energy levels produces polarized radiation in spectral lines which depends on the strength and orientation of the magnetic field (i.e., spectral line polarization induced by the Zeeman effect). However, unfortunately, the polarization induced by the Zeeman effect is of limited practical interest for the exploration of weak magnetic field in the upper solar atmosphere, because in the chromosphere and the transition region of the Sun the Zeeman splitting is typically smaller than the Doppler-broadened line width. Fortunately, scattering processes in the upper solar atmosphere produce linear polarization signals in Ly-alpha, which via the Hanle effect are sensitive to the weak magnetic fields (< 250 Gauss) expected for the solar transition region [1,2]. The Hanle effect can be defined as any modification of the atomic level polarization due to the presence of a magnetic field. The predicted line-center amplitude of the linear polarization signal caused by the Hanle effect in the upper solar chromosphere typically varies between 0.1% and 1%, approximately, depending on the strength and orientation of the magnetic field [2]. Therefore we need to develop and instrument capable of achieving a polarimetric sensitivity of at least 0.1% in the fractional linear polarization. The instrument layout of the CLASP is shown in Figure 1. The solar image is focused at the slit by the telescope mirrors. After the slit, a rotating half-waveplate modulates the polarization state, the polarization analyzer picks up a specific linear polarization state, the grating disperses the light, and finally the spectrograph cameras observes the spectra of Ly-alpha. As the polarimeter determines the polarization property of the CLASP, comprehensive tests of its performance before the launch are critical for the accurate measurement of the polarization.

---

[*] watanabe@kwasan.kyoto-u.ac.jp; phone 81 75 753 3893; fax 81 75 753 4280

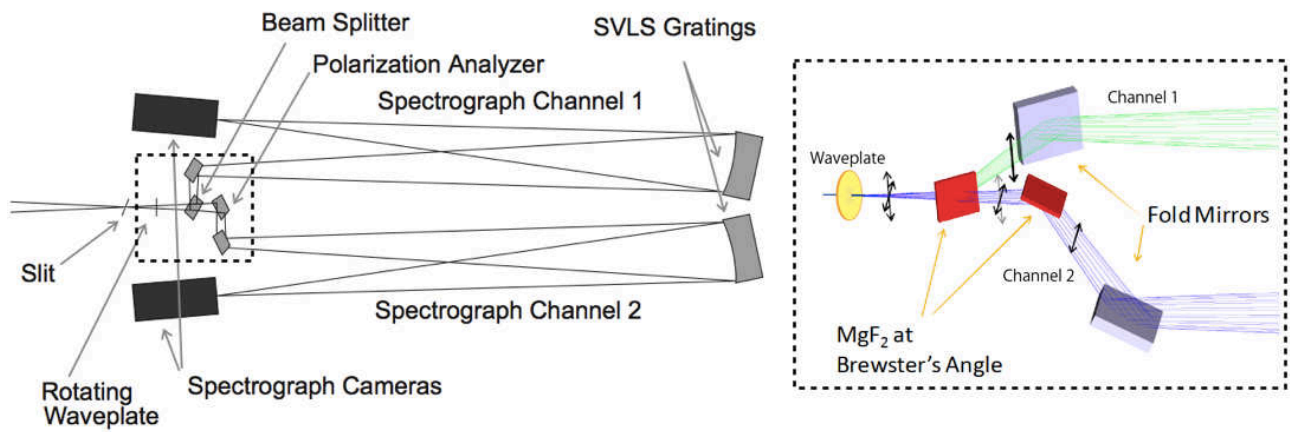

Figure 1. The instrument layout of CLASP after the slit (left) and the close-up to the polarimeter (right). The primary and secondary mirrors (before the slit, not shown) create the solar image focused at the slit. The polarimeter sits in the middle after the slit, consisting of a rotating half-waveplate, a beam splitter, a polarization analyzer, and two fold mirrors to send the light towards the gratings.

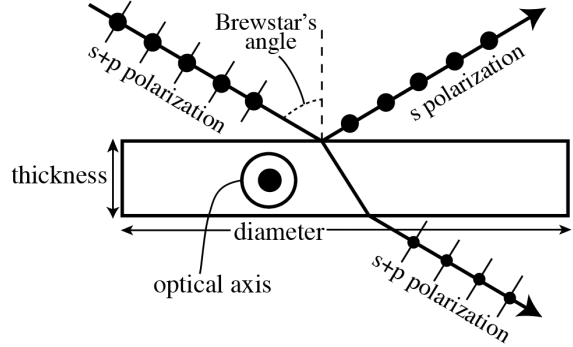

Figure 2. The concept of the MgF$_2$ beam splitter

The polarimeter of CLASP (Figure 1 right panel) is designed to detect linear polarization signals at the Ly-alpha wavelength. It consists of a half-waveplate, a beam splitter, and a polarization analyzer. We plan to use Magnesium Fluoride (MgF$_2$) as a material of a half-waveplate, a beam splitter, and a polarization analyzer. MgF$_2$ is a birefringent crystal with two axes with an ordinary refractive index $n_o$ and an extraordinary refractive index $n_e$ ("optical axis"), and offers high transparency at ultraviolet wavelength. The half-waveplate is located after the slit and rotates the direction of linear polarization. Both the beam splitter and the polarization analyzer are placed at the Brewster's Angle, but the polarization analyzer is rotated 90-degree around the incident beam relative to the beam splitter. The light with electric vector perpendicular to the plane of incidence is called s-polarization and the light with electric vector parallel to the plane of incidence is called p-polarization. The beam splitter (Figure 2) acts as a polarization analyzer for its reflected beam. The reflected beam ("channel 1") is composed entirely of s-polarized light. The transmitted beam through the beam splitter ("channel 2") is predominantly p-polarized but includes some s-polarized light. The reflection at the polarization analyzer eliminates the s-polarized component in the transmitted beam. Since MgF$_2$ is a birefringent material, transmission of the beam splitter has the potential to modify the polarization. This effect is eliminated by aligning the optical axis of the beam splitter to be parallel to the s-polarization direction.

The planned observing mode consists of rotating the waveplate at a constant rate and acquiring 16 exposures per waveplate rotation of 22.5-degree. We perform the demodulation of the observed signal afterwards on the ground. By observing the signals at two channels simultaneously, we can perform the cross-calibration between two channels, compensate for the change in source intensity or other instrumental polarizations.

As there has been no polarimeter for Ly-alpha in practical use, we initiated designing and developing the polarimeter components. The performance of the components is strongly dependent on the optical properties of MgF$_2$ at Ly-alpha. Therefore, comprehensive testing programs are performed using the light source at the Ultraviolet Synchrotron Orbital Radiation Facility (UVSOR) facility [3] at the Institute for Molecular Sciences. In the following sections, we first describe the setup in the UVSOR beamline in Section 2. In Section 3 we describe the measurement of reflectance and transmission ratio against the incident angle, and the retardation measurement of a zeroth-order waveplate. Depending on the optical constants derived from these measurements, the designs and specifications of a half-waveplate, a beam splitter, and a polarization analyzer are determined. By assembling the tested components, we constructed a prototype of the CLASP polarimeter, and performed simultaneous measurements of orthogonally linear polarizations at two channels (Section 4). Finally a summary is given in Section 5.

## 2. BEAMLINE SETUP

Our experiments were performed using the beamline BL7B [4,5] at the UVSOR facility. The light source at the UVSOR facility is generated by the synchrotron radiation with the storage ring of 50-m perimeter. The beamline BL7B consists of a normal incidence monochromator that covers the wavelength range of 40-1000 nm with a grating. In the wavelength range around Ly-alpha, the radiation intensity increases almost linearly as wavelength (5% increase per 1 nm). We can choose the widths of slits located at the entrance and the exit of the grating depending on the number of photons we need. Usually both slits are set at the same width. For example the width setting of 500 μm at both slits gives the beam with the photon count of $10^{10}$ photons s$^{-1}$ and the full-width-half-maximum (FWHM) bandwidth of 0.26 nm. The beam size at the focus position is about 2 mm Φ. The synchrotron radiation at the beamline BL7B is highly linearly polarized in the horizontal direction (≈ 90%). In addition, we installed one $MgF_2$ Brewster's reflection and one fold mirror in front of the experimental space, so that the input light is almost perfectly linearly polarized (> 99%) in the horizontal direction.

An accurate absolute wavelength calibration is important to characterize the optical properties of the polarimeter at Ly-alpha wavelength. The wavelength calibration was done by correlating the $O_2$ line spectra in the reference [6,7] around 121 nm at ≈ 0.3 torr vacuum level with the slits as narrow as possible. This method achieves 0.1 nm accuracy of the absolute wavelength calibration. The crosscheck is done by using the 10 nm bandwidth Ly-alpha filter supplied by the Acton Optics.

## 3. DESIGNS AND PERFORMANCES OF POLARIMETER COMPONENTS

In this section, we describe the design of the CLASP polarimeter components, i.e., a half-waveplate, a beam splitter, and a polarization analyzer. The design and specification depend on the optical constants of $MgF_2$ at Ly-alpha wavelength. To derive the optical constant, we performed two experiments: reflectance and transmittance measurement for s- and p-polarized light, and the retardation measurement of a zeroth-order waveplate. We first describe the two experiments, and then proceed to the design.

### 3.1 Reflectance and transmittance measurement

We performed the reflectance ($R$) and transmittance ($T$) measurement using the 2 mm thick $MgF_2$ plate for s- and p-polarization. From this measurement we can derive the optical constants of $MgF_2$, the value of the Brewster's Angle, and the throughput of $MgF_2$ plate. By measuring $R$ and $T$ as a function of incident angle, the ordinary refractive index $n_o$ and extinction coefficients along ordinary and extraordinary axes ($k_o$, $k_e$) are derived with a fitting using the Kramers-Krönig relation. The accurate measurement of the optical constants of the $MgF_2$ is important in its own because there have been only a few literatures of the optical constant measurement in the vacuum-ultraviolet [8,9]. The small difference between the ordinary and the extraordinary refractive index $n_e − n_o$ is more accurately derived by the retardation measurement described in Section 3.2, so we do not include $n_e$ into the fitting parameters.

The experiment was performed at UVSOR BL7B in March 2011. The wavelength was set to 121.567 nm with accuracy of 0.1 nm. The input light has the FWHM bandwidth of 0.16 nm. The input linearly polarized light is reflected by the 2 mm thick $MgF_2$ plate. The reflected light is detected by a 1-dimensional silicon photodiode with size of 15 mm × 15 mm. By setting $MgF_2$ plate aside we took the direct light for normalization of the reflection or transmission. A stepping motor with one pulse of 0.0025-degree controls the incident angle against the plate and the detector position. The detector position is fine-tuned so that it covers both the main beam and the ghost beam (internal reflected component). Because of the blocking by the plate holders, the measurable incident angle range is from 11-degree to 75-degree for $R$ and from 0-degree to 63-degree for $T$. The measurements of $R$ and $T$ are done at incident angles in the interval of 4-degree.

The measured $R$ and $T$ against the incident angle are shown in Figure 3. The signal-to-noise ratio of $R$ and $T$ is 20-100 originating from the fluctuation of the input light intensity. The dark current is subtracted, and the normalization by direct light is applied. The solid lines in Figure 3 are the result of the least-square fitting to the Kramers-Krönig relation. We take into account the first ghost and the second ghost, but neglected the third and higher ghosts into the fitting procedure. The intensity of the third ghost is typically 2% to the main beam. In the fitting assumptions, the result of the retardation measurement (Section 3.2) is used, i.e., $n_e = n_o + 0.00387$. As the output of the fitting, we got three parameters; $n_o = 1.664 \pm 0.003$, $k_o = (1.08 \pm 0.6) \times$ E−6, $k_e = (1.23 \pm 0.10) \times$ E−6. From these optical constants, the Brewster's Angle at Ly-alpha center is calculated to be $59.00 \pm 0.05$ degree, and the ghost contributions at the Brewster's Angle are 37% ($R$p), 26% ($R$s), 0% ($T$p), 3% ($T$s) to the main beam.

The performance of the polarization analyzer is expressed by the polarization extinction ratio (PER:= $R$p/$R$s) at the Brewster's Angle. As $R$p = 0.23% and $R$s = 29% at 59-degree incident angle (ghosts included), PER is 0.0079. The 2 mm thick $MgF_2$ plate transmits 77.8% of the p-polarized light. The PER of the polarizing analyzer without the ghost contributions is 0.0066.

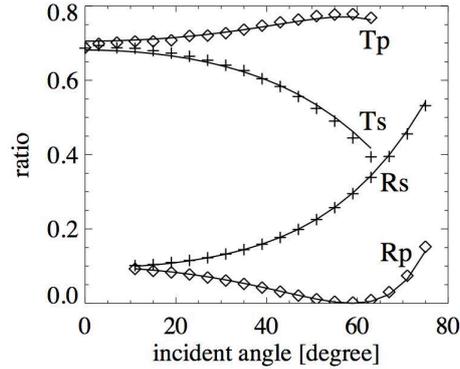

Figure 3. The measured $R$ and $T$ against the incident angle. The subscript "p" and "s" stands for the p-polarized and the s-polarized light, respectively. The solid lines are the fitting results based on the Kramers-Krönig relation.

## 3.2 Retardation measurement

The prototype waveplates are custom made by Kogaku Giken Corporation to the specified thickness difference. The total thickness of the waveplate is 1 mm and the diameter is 20 mm. The optical axes of two stacked plates are rotated by 90-degree within 0.2-degree accuracy. We measure the retardation of the waveplate prototypes several times, revising the thickness difference and fabricating a new prototype each time to attain a retardation of 180-degree as accurately as possible. In this section we report the retardation measurement of a zeroth-order waveplate with the thickness difference of $14.57 \pm 0.02$ μm. The experiment was performed at UVSOR BL7B in March 2011. From this measurement, we can derive the small difference of $n_e$ and $n_o$ against wavelength, leading to the thickness difference needed for a half-waveplate at Ly-alpha center.

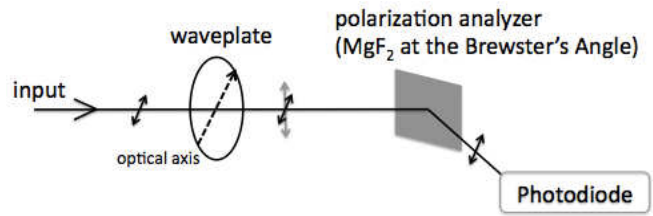

Figure 4. The configuration of the retardation measurement.

The configuration of the retardation measurement is illustrated in Figure 4. The input beam (linear polarization in horizontal direction) first goes through the waveplate in normal incidence. Then the light is reflected by a Brewster's reflection (59-degree rotated around the vertical direction) of an $MgF_2$ plate to take out the p-polarized light. One method to derive the retardation ($\delta$) of a waveplate is observing the modulation through a rotating waveplate and a polarization analyzer by input a known polarization at a specific wavelength. When the waveplate's optical axis is parallel to the direction of input beam's polarization, no rotation of the input linear polarization is given and we obtain the maximum signal ($I_{max} = I_{input}$ * (waveplate transmittance) * $Rs$(59-degree)). The signal obtained when the optical axis of the waveplate rotated by 45-degree becomes the minimum of the modulation, because the input light is rotated 90-degree after the waveplate and it corresponds to p-polarization to the polarization analyzer ($I_{min} = I_{input}$ * (waveplate transmittance) * $Rp$(59-degree)). The modulation amplitude ($I_{min} / I_{max}$) is expressed by $(1+\cos\delta)/2$ when the polarization analyzer is perfect (i.e., PER=0). In a general case (i.e., PER≠0) the modulation amplitude is expressed like this:

$$(\text{Modulation amplitude}) = [(1+\text{PER}) + (1-\text{PER}) \cos\delta] / 2 \quad (1)$$

The derivation of $\delta$ depends on PER, whose accurate value is difficult to know because of the small $Rp$ around the Brewster's angle. Also the dependency of the modulation amplitude to $\delta$ is almost flat around $\delta$=180-degree. This method is therefore not suitable for the retardation derivation around 180-degree. To overcome this problem, we took another method that takes advantage of the wavelength dependency of retardation. The retardation and $n_e - n_o$ value are related by the equation:

$$\delta = 2\pi |d_1 - d_2| (n_e - n_o) / \lambda \quad (\lambda : \text{wavelength}) \quad (2)$$

Here $d_1$ and $d_2$ stand for the thickness of the stacked two $MgF_2$ plates. The equation (2) tells that the retardation of a waveplate changes rapidly against wavelength, and so does the modulation amplitude. We take the position where the modulation amplitude becomes the minimum during the wavelength scan. The wavelength position to have 180-degree retardation is determined independently of PER.

Our experiment has a configuration similar to that shown in Figure 4, with the optical axis fixed at 45-degree ($I_{min}$) or 0-degree ($I_{max}$) around the input beam. The reflected light is detected by a 1-dimensional silicon photodiode with size of 15 mm × 15 mm. We scan the wavelength of the input light from 120.1 nm to 124.1 nm with the interval of 0.02 nm. The input light has a wavelength FWHM of 0.26 nm. As our absolute wavelength accuracy is 0.1 nm and the measured wavelength interval is 0.02 nm, the inclusive wavelength accuracy in this measurement is at most 0.12 nm. By dividing the two cases ($I_{min}$ devided by $I_{max}$), we get the modulation amplitude in the wavelength range from 120.1 nm and 124.1 nm. The dark current (typically 0.5% of the signal) is subtracted.

The result is plotted in Figure 5. The signal-to-noise ratio originated from the input beam stability is about 100. The modulation amplitude takes the minimum value of 0.012 (i.e., PER = 0.012) at 121.74 ± 0.12 nm. This is the wavelength where the 14.57 μm waveplate have 180-degree retardation. In Figure 5 we overplot the retardation derived from the equation (1) assuming PER = 0.012 throughout the wavelength range. In Section 3.1, we derived PER = 0.0079 at 121.567nm. The difference of PER = 0.012 and 0.0079 makes the retardation error of 1.8-degree. The retardation error coming from the wavelength determination accuracy (0.12 nm) is much larger (about 10-degree). The assumption of constant PER is negligible when the modulation amplitude is large because PER << 1.

Using the equation (2), we found that $n_e - n_o$ = 0.00418 at 121.74 ± 0.12 nm where $\delta$=180-degree, and $n_e - n_o$ = 0.00387 ± 0.00023 at 121.567 nm.

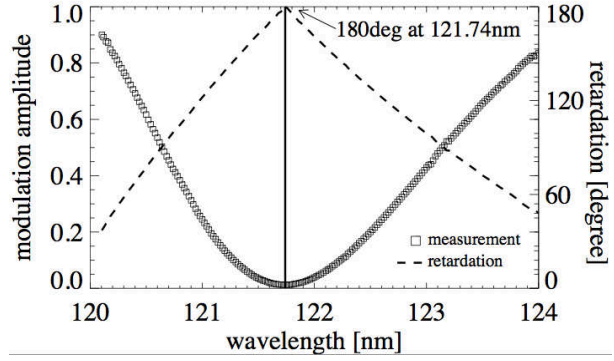

Figure 5. The modulation amplitude against the wavelength through the 14.57 um waveplate (plus symbol). A dash line shows the calculated retardation from equation (1).

**3.3 Design of the half-waveplate**
The waveplate is a zeroth-order design consisting of two stacked $MgF_2$ plates with slightly different thicknesses and their optical axes rotated by 90-degree from each other. The retardation of the waveplate is expressed by the equation (2). Using the derived $n_e - n_o$ value in Section 3.2 ($n_e - n_o$ = 0.00387 ± 0.00023 at 121.567 nm), we conclude that the waveplate with thickness difference of 15.71 um will work as a half-waveplate at Ly-alpha wavelength with the retardation error of 10-degree. The 10-degree retardation error originates mainly from the wavelength determination accuracy, and other elements such as the tolerance of the waveplate thickness difference are small enough to be neglected. The 10-degree retardation error is within tolerance of the CLASP polarimeter if we assume the observed maximum linear polarization is 1% and the observed maximum circular polarization is 0.1%.

Kogaku Giken Corporation supplies the flight waveplate as the prototype. The transmittance of the 1-mm thick waveplate at Ly-alpha is 85% from the reflectance and transmittance measurement in Section 3.1.

**3.4 Design of the beam splitter and the polarization analyzer**
In Section 3.1 we confirmed that the Brewster's reflection (59-degree) of an $MgF_2$ plate becomes a good polarization analyzer. The thickness of the beam splitter is 2 mm in order to secure enough amount of transmitted light. The 2-mm beam splitter transmits 77.8% of the p-polarized light. On the other hand, the thickness of the polarization analyzer is 15 mm to eliminate the ghost image. The throughputs of the two channels by the beam splitter and the polarization analyzer are 29% for channel 1 and 17% for channel 2.

The 2-mm thick beam splitter is demanded from the viewpoint of the throughput. One potential issue is the ghost image created by the 2-mm thick beam splitter, but we plan to mitigate this issue by ensuring that the plate is sufficiently thick, so that the ghost image is displaced far from the main spectral line. With a 2 mm thick $MgF_2$ plate, the ghost image is placed 0.32 nm in dispersion offset from the main spectral line on the detector, which is sufficient to avoid contamination of the spectral line.

## 4. PROTOTYPE OF THE CLASP POLARIMETER
We constructed a prototype of the CLASP polarimeter (Figure 6) by assembling the components whose performance evaluations are done in Section 3.1 and 3.2. The fold mirrors are not used. Instead of the fold mirrors, the detector holders are located at the exit of both channels. A CCD camera or a silicon photodiode can be installed at these detector holders. In this section we focus on the results detected with the two photodiodes. By installing two photodiodes for both channels, we can take the signal at both channels simultaneously at the same configuration, similar to the flight situation.

The experiment is performed at the UVSOR BL7B in March 2011.

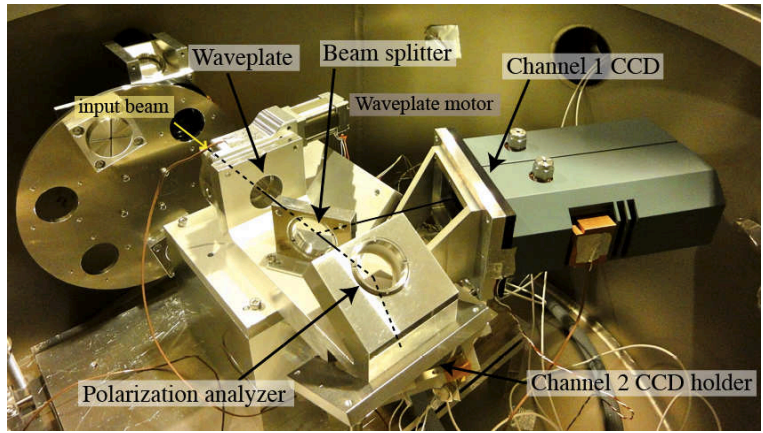

Figure 6. The photo of the prototype of the CLASP polarimeter

**4.1 Design of the prototype polarimeter**
The waveplate is a zeroth-order waveplate with 14.57 μm thickness difference, which is the same one tested in Section 3.1. The beam splitter is a 2 mm thick $MgF_2$ plate that we tested in Section 3.2. The polarization analyzer is an $MgF_2$ plate with 15 mm thickness, so as to secure the ghost elimination. The perimeters of the components and the size of the silicon detectors are large enough to cover the whole beam.

The waveplate rotation is controlled by a stepping motor with one pulse of 0.0025-degree. The detector positions are located roughly at the focus position of BL7B. By using a pair of autocollimated theodolites, we confirmed that both the beam splitter and the polarization analyzer were mounted at 59-degree to the beam within 3 arcmin accuracy. The normal incidence to the waveplate along the rotation axis is kept within 3 arcmin accuracy.

**4.2 Performance test of the prototype polarimeter**
The performance of the prototype polarimeter is verified at the wavelength of 121.74 nm, instead of Ly-alpha. At this wavelength, our waveplate has a retardation of 180-degree. The input light has the FWHM bandpass of 0.26 nm. We measured the modulation of the signal during one rotation of the waveplate at 4-degree interval. Figure 7 shows the signals measured in current (dark is subtracted). The detector installed for channel 2 has the sensitivity of 0.8 times that of the detector for channel 1.

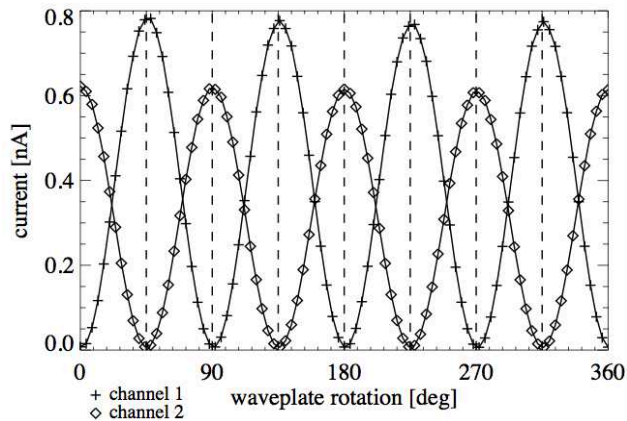

Figure 7. The modulation of the signal detected at channel 1 (+) and channel 2 (◇) versus waveplate rotation angle. The solid lines connects the simbols.

The observed modulation can be well fitted by the cos4θ (θ: waveplate rotation angle) function, and the phase difference between the two channel is 45 ± 1-degree (1-degree error is acceptable). This is consistent with the theoretical prediction, based on the assumption that the input light consists of perfect horizontal polarized light and the waveplate has a retardation of 180-degree. The minimums of the modulations are close to zero (≈ 0.006 nA for both channels in Figure 7), which demonstrates the good performance of the $MgF_2$ plate as a polarization analyzer.

In the actual observation, we assume a polarimeter model and compare the observed modulation with the predicted modulation based on our polarimeter model. Thus the demodulation from the observed signal into the polarization state of the input light depends on the accurate polarimeter modeling. For test, we calculated a virtual polarimeter model consisting of a half-waveplate and a perfect polarization analyzer. This virtual polarimeter model and the one constructed from the measurement matches within 2% scale error. This 2% scale error is good enough for the CLASP observation, because the observed linear polarization in the solar atmosphere is of the order of 0.1%, and thus the 2% scale error makes the linear polarization error of 0.1% × 2% = 0.002%, which is below the photon noise.

We also performed the experiment using the circular polarized light as input by using a waveplate with thickness difference of 7.04 um (δ=87-degree at 121.74 nm). If the waveplate works as half waveplate, the polarimeter prototype has no response to the circular polarization, i.e., no modulation corresponding to circular polarization will be observed. Indeed the measurement supports this prediction. It proves that the crosstalk from circular polarization is very small.

## 5. SUMMARY

In this paper we report the development of the CLASP polarimeter. We design the CLASP polarimeter to be optimized for measuring the weak linear polarization signal in the solar chromosphere and the transition region created by the Hanle effect. The half-waveplate rotates the direction of linear polarization, and the beam splitter is installed so that it allows cross-calibration of the two channels. Before stepping into the polarimeter design, we performed the basic measurements of $MgF_2$ optical constants and retardations of a zeroth-order waveplate. The reflection of $MgF_2$ plate at the Brewster's Angle is confirmed to work as a good polarization analyzer for Ly-alpha. The zeroth-order half-waveplate is possible to fabricate with the retardation error of 10-degree. By assembling the evaluated components, the CLASP polarimeter prototype is constructed. The test using the polarimeter prototype succeeded in measuring the orthogonal linear polarization signals at two channels simultaneously. The crosstalk from circular polarization is confirmed to be negligible.

At this moment, there are no critical issues about the polarimeter that exceed the scientific tolerances. Our development of the polarimeter used in the vacuum-ultraviolet opens a new method for the space polarimetry.

## REFERENCES

[1] Trujillo Bueno, J., "Modeling Scattering Polarization for Probing Solar Magnetism", in Solar Polarization 6, eds. J. Kuhn et al., ASP Conf. Series Vol. 437, 83 (2011).
[2] Trujillo Bueno, J., Stepan, J., Casini, R., "The Hanle Effect of the Hydrogen Ly-alpha Line for Probing the Magnetism of the Solar Transition Region", ApJ Letters, in press (2011).
[3] www.uvsor.ims.ac.jp/defaultE.html
[4] Fukui, K., Nakagawa, H., Shimoyama, I., Nakagawa, K., Okamura, H., Nanba, T., Hasumoto, M. and Kinoshita, T., "Reconstruction of BL7B for UV, VIS and IR spectroscopy with a 3 m normal-incidence monochromator", J. Synchrotron Rad., 5, 836-838 (1998).

[5] Fukui, K., Miura, H., Nakagawa, H., Shimoyama, I., Nakagawa, K., Okumura, H., Nanba, T., Hasumoto, M. and Kinoshita, T., "Performance of IR-VUV normal incidence monochromator beamline at UVSOR", Nuclear Instruments and Methods in Physics Research A, 467-468, 601-604 (2001).
[6] Mason, N. J., Dawes, A., Holtom, P. D., Mukerji, R. J., Davis, M. P., Sivaraman, B., Kaiser, R. I., Hoffmann, S. V. and Shaw, D. A., "VUV spectroscopy and photo-processing of astrochemical ices: an experimental study", Faraday Discuss., 133, 311-329 (2006).
[7] Krupenie, P. H., "The Spectrum of Molecular Oxygen", J. Phys. Chem. Ref. Data, 1, 423-534 (1972).
[8] Williams, M. W. and Arakawa, E. T., "Optical properties of crystalline $MgF_2$ from 115 nm to 400 nm", Appl. Opt. 18, 1477-1478 (1979).
[9] Laporte, P., and Subtil, J. L., Courbon, M., Bon, M. and Vincent, L., "Vacuum-ultraviolet refractive index of LiF and $MgF_2$ in the temperature range 80–300 K", J. Opt. Soc. Am., 8, 1062-1069 (1983).